# Irreversibility line and low-field grain-boundary pinning in electron-doped superconducting thin films

S. Sergeenkov

*Grupo de Materiais e Dispositivos, Centro Multidisciplinar para o Desenvolvimento de Materiais Cerâmicos - CMDMC, Departamento de Física, Universidade Federal de São Carlos - UFSCar, Caixa Postal 676, CEP 13565-905, São Carlos - São Paulo, Brasil and Bogoliubov Laboratory of Theoretical Physics, Joint Institute for Nuclear Research, 141980 Dubna, Moscow Region, Russia*

A.J.C. Lanfredi and F.M. Araujo-Moreira

*Grupo de Materiais e Dispositivos, Centro Multidisciplinar para o Desenvolvimento de Materiais Cerâmicos - CMDMC, Departamento de Física, Universidade Federal de São Carlos - UFSCar, Caixa Postal 676, CEP 13565-905, São Carlos - São Paulo, Brasil*

**Abstract**

AC magnetic susceptibilities of electron-doped $Pr_{1.85}Ce_{0.15}CuO_4$ (PCCO) and $Sm_{1.85}Ce_{0.15}CuO_4$ (SCCO) granular thin films have been measured as a function of temperature and magnetic-field strength. Depending on the level of homogeneity of our films, two different types of the irreversibility line (IL), $T_{irr} \equiv T_p(H)$, defined as the intergrain-loss peak temperature in the imaginary part of susceptibility and obeying the law $1 - T_p/T_C \propto H^q$, have been found. Namely, more homogeneous PCCO films (with grain size of the order of $2\mu m$) are best-fitted with $q=2/3$ while less homogeneous SCCO films (with grain size of the order of 500 nm) follow the IL law with $q=1$. The obtained results are described via the critical-state model taking into account the low-field grain-boundary pinning. The extracted pinning-force densities in more granular SCCO films turn out to be four times larger than their counterparts in less granular PCCO films.

*Packs: 74.25.Ha; 74.25.Qt; 74.78.Bz; 74.81.-g*

## 1. Introduction

The measurement of AC magnetic susceptibility still remains one of the most powerful methods to obtain important information on dissipation mechanisms in high-$T_C$ superconductors (HTS). To get useful information from such experiments, however, very careful control of sample's microstructure is required. While in high enough magnetic fields the dissipation is known to be dominated by flux motion of Abrikosov vortices [1-4], the low-field dissipation mechanisms (especially, in inhomogeneous and granular superconductors) are less obvious due to the numerous grain-boundary related effects which are better treated by the Josephson physics [5-7].

Here we present a comparative study of low-field AC magnetic susceptibility data on more homogeneous (with grain size of the order of 2μm) $Pr_{1.85}Ce_{0.15}CuO_4$ (PCCO) and less homogeneous (with grain size of the order of 500 nm) $Sm_{1.85}Ce_{0.15}CuO_4$ (SCCO) thin films. The main idea of the experiments here reported is to study the influence of inhomogeneity on the dissipative properties of electron-doped thin films via the behavior of the irreversibility line (IL), $T_{irr} \equiv T_p(H)$, defined as the intergrain-loss peak temperature in the imaginary part of susceptibility $\chi''(T,H)$. This influence was found to result in a much higher pinning ability of less homogeneous SCCO thin films obeying the IL law $1 - T_p/T_C \propto H^q$ with q=1 as compared to more homogeneous PCCO films with flux-creep-mediated exponent q=2/3.

## 2. Samples Characterization and Experimental Procedure

A few PCCO and SCCO thin films (d=200nm thick) grown by pulsed laser deposition on standard $LaAlO_3$ substrates were used in our measurements (for discussion on different preparation techniques and chemical phase diagrams of electron-doped superconducting materials, including polycrystalline samples, single crystals, and thin films, see, e.g., [8-12] and further references therein). All samples showed similar and reproducible results. The structural quality of our samples was verified through both x-ray diffraction (XRD) and scanning electron microscopy (SEM) together with energy dispersive spectroscopy (EDS) technique. The SEM experiments reveal that PCCO films are of higher structural quality (more homogeneous) than SCCO films which show a pronounced granular structure (see Figure 1).

The average grain size in typical PCCO and SCCO films is estimated to be around 2μm and 0.5μm, respectively. Measurements of the real ($\chi'$) and imaginary ($\chi''$) parts of AC susceptibility were performed by using a MPMS magnetometer from the Quantum Design equipped with AC modulus [13-16]. All data are chosen from samples with the same dimensions and well placed parallel to the field in order to decrease the demagnetization correction. The symbol size used for data presentation takes into account error bars based on the temperature stability, reproducibility, and system precision. To account for a possible magnetic response from substrate, we measured several stand alone pieces of the substrate. No tangible contribution due to magnetic impurities was found.

## 3. Results and Discussion

A typical temperature behavior of the measured complex AC magnetic susceptibility in PCCO and SCCO films in a small magnetic field (of amplitude $h_{AC}$=1.0Oe and frequency $f_{AC}$ =1.0kHz) is shown in Fig.2. The field dependence of the imaginary part $\chi''$ of the AC susceptibility in both films for the temperatures close to $T_C$ is depicted in Fig.3.

Due to small values of the applied magnetic field, it is natural to associate the peak temperatures $T_p(H)$ in Fig.3 with *intergrain* losses. The extracted values of the irreversibility temperature $T_p(H)$ for both samples are shown in Fig.4 in the form of the log-log plots. As is seen, more homogeneous PCCO films are well-fitted by the flux-creep mediated IL obeying the law $1- T_p/T_C \propto H^q$ with q=2/3 while less homogeneous SCCO films (with grain size of the order of 500 nm) follow the IL law with q=1.

To interpret the above findings, we follow Müller's approach [17] (based on the Kim-Anderson critical-state model [18]) according to which the low-field dependence of the IL temperature $T_p(H)$ is governed by the following implicit equation (hereafter $H \equiv h_{AC}$)

$$\left[1+\frac{H}{H_C(T_p)}\right]^2 = 1+\frac{2d\mu_J(T_p)}{\mu_0 \mu_{eff}(T_p)[H_C(T_p)]^2} \qquad (1)$$

where

$$\mu_{eff}(T) = \frac{2I_1(R/\lambda)\lambda}{I_0(R/\lambda)R} \qquad (2)$$

Here, $\lambda(T)$ is the London penetration depth, R is the average grain size, d is the film thickness, $\mu_{eff}(T)$ is the effective permeability of granular film, $H_C(T)$ is the characteristic field (see below), $\mu_J(T)$ is the so-called pinning-force density, and $I_0$ and $I_1$ are modified Bessel functions of the first kind. Notice that Eq.(1) is valid for applied fields larger than the lower Josephson field $H_C(T) = \frac{\phi_0}{4\pi\mu_0 \lambda(T)R}$ when vortices nucleate along grain boundaries. These intergranular Josephson vortices are imbedded into a diamagnetic medium with effective permeability $\mu_{eff}(T)$ whose temperature dependence, in view of Eq.(2), is governed by the London penetration depth $\lambda(T) = \frac{\lambda(0)}{\sqrt{1-T/T_C}}$. The observed difference in behavior of IL is attributed to difference in average grain sizes in PCCO and SCCO films which, according to SEM scans (see Figure 1) are around R=2μm and R=500 nm, respectively.

Taking into account the explicit temperature dependence of the pinning-force density within the grain-boundary pinning model [19]

$$\mu_J(T) = \mu_J(0)(1-T/T_C)^{3/2} \qquad (3)$$

we propose the following scenario for the observed IL behavior.

Since near $T_p$ in more homogeneous PCCO films (see Fig.1a) $R > \lambda(T)$, and hence $\mu_{eff}(T) \approx 2\lambda(T)/R$, from Eq.(1) we find the usual flux-creep dominated law (see Fig.4a)

$$1 - \frac{T_p}{T_C} = AH^{2/3} \qquad (4)$$

with $A = \left[\dfrac{2\mu_0 \lambda(0) H_C(0)}{\mu_J(0) dR}\right]^{2/3}$.

On the other hand, in more granular SCCO films (see Fig.1b) near $T_p$ we have the opposite situation since in this case $R < \lambda(T)$, and hence $\mu_{eff}(T) \approx 1$. As a result, Eqs.(1)-(3) bring about the observed linear behavior of the IL (see Fig.4b)

$$1 - \frac{T_p}{T_C} = BH \qquad (5)$$

with $B = \dfrac{\mu_0 H_C(0)}{\mu_J(0) d}$.

By calculating the coefficients A and B from the IL curve slopes on a log-log plot, we can estimate the pinning-force densities $\mu_J(0)$ for both materials. Using for the film thickness d=200nm, London penetration depths [20] $\lambda_P(0)$=250nm, $\lambda_S(0)$=500nm, and average grain sizes R=2µm and R=0.5µm, from Eqs.(4) and (5) we obtain $\mu_{JP}(0)$=3x10$^4$TA/m$^2$ and $\mu_{JS}(0)$= 1.2x10$^5$TA/m$^2$ for the pinning-force densities of PCCO and SCCO films, respectively. As expected, the above pinning values are larger than those seen in bulk granular materials [5-7]. Thus, for small applied magnetic fields, the flux pinning is dominated by the so-called electromagnetic pinning scenario characterized by the London pentration depth rather than coherence length (the latter is responsible for the so-called core pinning scenario in high enough magnetic fields). Within this scenario, the observed higher pinning ability of SCCO films near $T_p$ can be attributed to a perfect match between the average grain size R and the correspondent London penetration depth $\lambda_S(T_p)$. While in the case of a more homogeneous PCCO film (with the average grain size of R=2000nm) the ratio $\lambda_P(T_p)/R$ is much less optimal leading to a lower pinning ability of these films. And finally, it is instructive to point out that the obtained here results on low-field irreversibility lines in our granular films (governed by grain-boundary pinning of coreless Josephson vortices) principally differ from the high-field irreversibility lines observed in electron-doped single crystals (dominated by core pinning of Abrikosov vortices, including particular scenarios for melting of the vortex lattice) [4].

## 4. Conclusion

In summary, by analyzing the measured AC magnetic susceptibilities of electron-doped $Pr_{1.85}Ce_{0.15}CuO_4$ (PCCO) and $Sm_{1.85}Ce_{0.15}CuO_4$ (SCCO) thin films as a function of temperature and magnetic-field strength, we associated the irreversibility line (IL), $T_p(H)$, obeying the law $1- T_p/T_C \propto H^q$, with the intergranular peaks in the imaginary part of AC susceptibilities. We found that more homogeneous PCCO films (with grain size larger than 2μm) are best-fitted with q=2/3 while less homogeneous SCCO films (with grain size less than 500 nm) follow the IL law with q=1. The obtained results are described in the framework of the Kim-Anderson critical-state model taking into account the grain-boundary pinning of Josephson vortices. The extracted pinning-force densities in granular SCCO films found to be four times larger than their counterparts in more homogeneous PCCO films.

**Acknowledgements**

We gratefully acknowledge financial support from Brazilian agency FAPESP. We also thank S. Anlage, C. J. Lobb and R. L. Greene from the *Center for Superconductivity Research* (University of Maryland at College Park) for useful comments and discussions.

## FIGURES

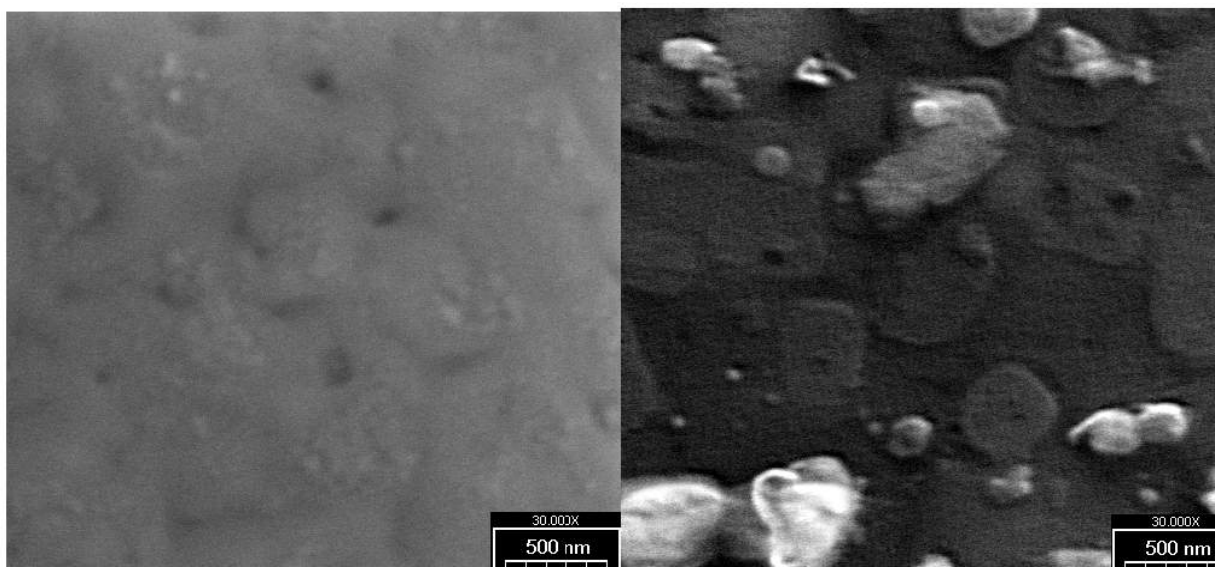

**Figure 1**. SEM scan photography of PCCO (left) and SCCO (right) samples (magnification 30000 times).

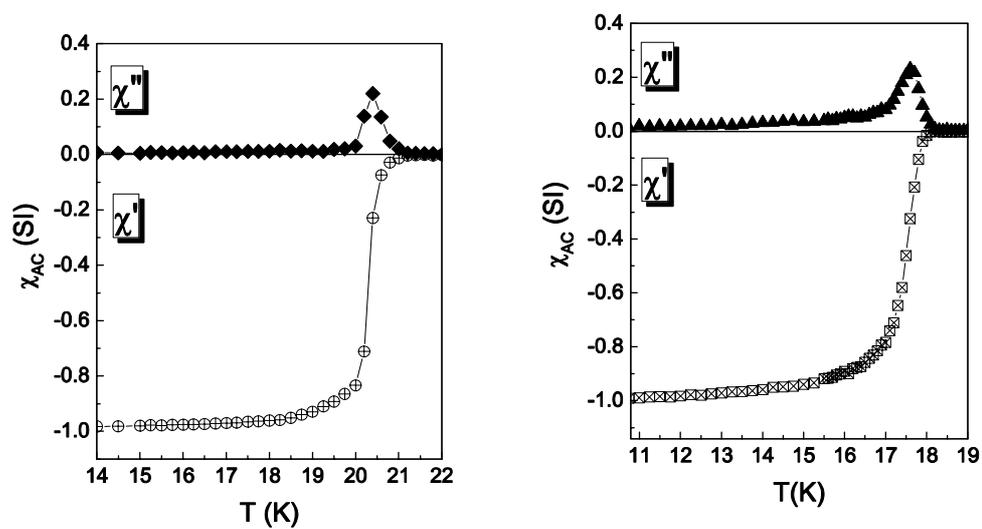

**Figure 2.** Temperature behavior of the AC susceptibility measured on PCCO (left) and SCCO (right) thin films for magnetic field of amplitude $h_{AC}$=1.0 Oe and frequency $f_{AC}$ =1.0 kHz.

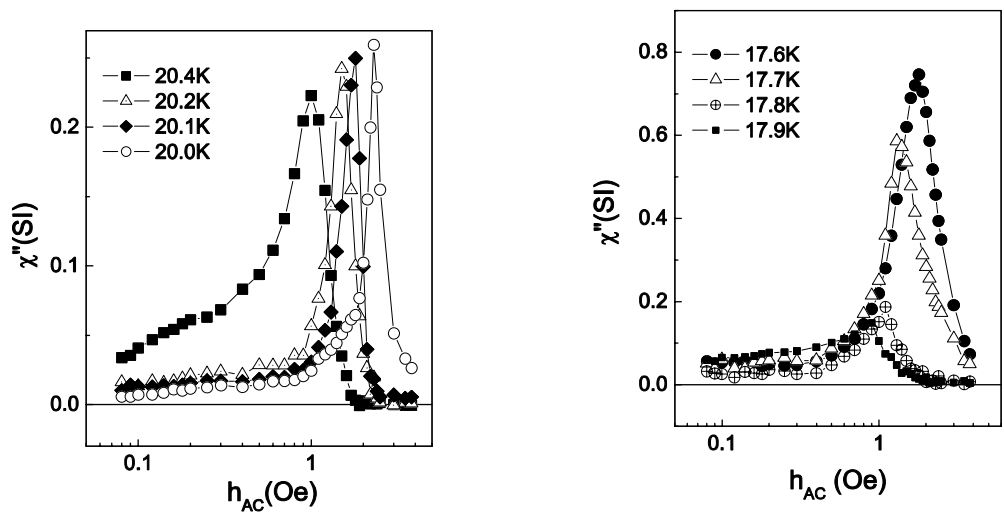

**Figure 3.** Magnetic field behavior of the imaginary part of AC susceptibility measured on superconducting thin films at different temperatures: PCCO (left) and SCCO (right).

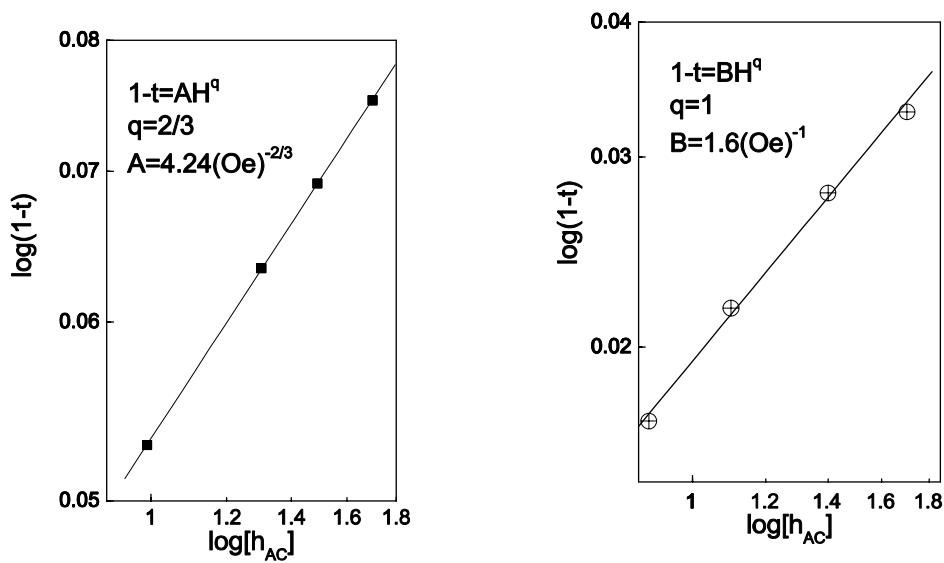

**Figure 4.** Log-log plot of the irreversibility lines $1-t \propto H^q$ (extracted from AC susceptibility data shown in Figure 3) for PCCO (left) and SCCO (right) films. Solid lines are the best fits according to Eqs.(1)-(5).